\documentclass[twocolumn,superscriptaddress,showpacs,aps]{revtex4-1}
\usepackage{ulem,bm}
\usepackage{amsmath, upgreek, amssymb, graphicx, paralist, gensymb,epstopdf}
\usepackage[colorlinks, linkcolor=blue, citecolor=blue, urlcolor=blue]{hyperref}
\usepackage{array,epsfig,amsmath,amssymb}
\usepackage{graphicx} 
\usepackage{dsfont}
\usepackage{color}

\def\bra#1{\langle #1|}
\def\ket#1{|#1 \rangle}
\def\bracket#1#2{\langle #1|#2 \rangle}

\begin{document}

\title{Optimal quantum phase estimation with \\
generalized multi-component Schr\"odinger cat states}

\author{Seung-Woo Lee}
\email{swleego@gmail.com}
\affiliation{Quantum Universe Center, Korea Institute for Advanced Study, Seoul 02455, Korea}

\author{Su-Yong Lee}
\email{papercrane79@gmail.com}
\affiliation{Quantum Physics Technology Directorate, Agency for Defense Development, Daejeon 34186, Korea}

\author{Jaewan Kim}
\affiliation{School of Computational Sciences, Korea Institute for Advanced Study, Seoul 02455, Korea}


\begin{abstract}
In this paper, we are interested in detecting the presence of a nearby phase-sensitive object, where traveling light works out under a low-photon loss rate. Here we investigate the optimal quantum phase estimation with generalized multi-component Schr\"odinger cat states. In addition, we show the optimal conditions of the generalized multi-component cat states for the phase estimation in a lossless scenario. We then demonstrate that the generalized multi-component cat states can beat the performances of the NOON and two-mode squeezed vacuum states in the presence of small loss, while maintaining the quantum advantage over the standard quantum limit, attainable by coherent states. Finally, we propose a generation scheme of the entangled multi-component cat states with current or near-term optical technologies.
\end{abstract}

\maketitle



\section{Introduction}
Quantum parameter estimation is a key theory of quantum metrology \cite{Hel}. By measuring a probe which interacted with a parameter, we can obtain the information on the parameter statistically. If the mean value of the parameter is located on the true value of it, then we only need to focus on reducing the width of the probability distributions, i.e., the root-mean-square error (RMSE). Under the constraint of input energy $N$, it is the best strategy of reducing the RMSE as much as we can. For classical input states, the RMSE of a parameter is lower bounded by a scaling of $1/\sqrt{N}$ which is called the standard quantum limit (SQL) achieved with coherent states. For quantum input states, it is lower bounded by a scaling of $1/N$ which is called the Heisenberg limit (HL) achieved with entangled or squeezed states \cite{Dow}.

We are interested in a physical parameter, i.e., a phase that can be set with a phase shifter in an interferometry. Since the phase shifter produces a path length difference in the interferometry, we can infer the phase parameter by measuring an output signal which went through the path length difference \cite{Dow}.
For example, if we inject a coherent state into an interferometer which does not include a phase shifter in any arm, there is no click event in one of the output modes. Once we include the phase shifter in one arm, then there will be a possibility of having a click event in the other output mode. 
 In a laboratory, the phase parameter can play a role of gravitational waves that produced a path length difference in a huge interferometer \cite{LIGO}.

For a probe state, we choose the generalized multi-component cat states defined as the equally superposed coherent states on a circle in phase space. The characteristics of multi-component cat states were first studied \cite{psp1,psp2} and their applications to universal quantum computation with error corrections \cite{Jkim2015,Bergmann16,Grimsmo20} as well as quantum cryptography \cite{SWL19} have been actively studied recently. Generation of multi-component cat states has been proposed and reported in cavity quantum electrodynamics (QED) \cite{cavity}, circuit QED \cite{cQED,cQED1,Wang16}, optomechanical system \cite{opto}, and traveling optical systems \cite{SWL19}. Quantum phase estimation using entangled coherent states was first studied in Ref.~\cite{Joo11}, and the analysis was further expanded to a multi-component cat state by one of us \cite{SYLee15}. However, it was not revealed about whether the multi-component cat state was set by its optimal conditions. Thus, it may be of importance to check the optimal conditions of multi-component cat state for the application to the quantum phase estimation protocol not only in lossless scenario but also in the presence of losses.
Here we consider generalized multi-component cat states to estimate a phase parameter in lossy interferometry, where the photon loss process can occur in both arms. In terms of quantum Fisher information (QFI) characterizing the ultimate precision, we show that the multi-component cat states can surpass the performance of the NOON and two-mode squeezed vacuum (TMSV) state as well as beat the SQL. The idea can be applied to a proximity sensor.

This paper is organized as follows. We start with the brief introduction of the generalized multi-component cat state in Sec.~\ref{sec:GCS}. In Sec.~\ref{sec:OPS}, we investigate the phase estimation by entangled multi-component cat states. The optimal conditions for the phase estimation in a lossless scenario is presented. We analyze the effect of losses in Sec.~\ref{sec:LOSS}. It is demonstrated that multi-component cat states can outperform the NOON and TMSV states under small losses. Finally, in Sec.~\ref{sec:GS}, we propose a scheme for producing the entangled multi-component cat states based on cross-phase modulators. We conclude our work in Sec.~\ref{sec:CON}.

\section{Generalized cat states} 
\label{sec:GCS}

We begin with the definition of the generalized multi-component cat states. Consider a superposition of the coherent states $\{|\alpha\rangle, |\alpha\omega\rangle,...,|\alpha\omega^{d-1}\rangle\}$ which are equally distributed on a circle in phase space, where $\alpha$ is the amplitude and $\omega=\exp{(2\pi i/d)}$ with a positive integer $d$. By setting different relative phases of the coherent states, we can define the {\em generalized multi-component cat states},
\begin{equation}
\label{eq:pns1} 
\ket{C_{d,k}(\alpha)}\equiv\frac{1}{\sqrt{{\cal M}_{d,k}(\alpha)}}\sum^{d-1}_{q=0}\omega^{-kq}\ket{\alpha\omega^{q}},
\end{equation}
where $k\in\{0,1,...,d-1\}$ determines the relative phases among the different coherent states. For example, when $d=2$, $\ket{C_{2,0}(\alpha)}\propto \ket{\alpha}+\ket{-\alpha}$ and $\ket{C_{2,1}(\alpha)}\propto \ket{\alpha}-\ket{-\alpha}$ are the (well-known) even and odd cat states, respectively. When $d=4$, $\ket{C_{4,0}(\alpha)}\propto \ket{\alpha}+\ket{i\alpha}+\ket{-\alpha}+\ket{-i\alpha}$, $\ket{C_{4,1}(\alpha)}\propto \ket{\alpha}-i\ket{i\alpha}-\ket{-\alpha}+i\ket{-i\alpha}$, $\ket{C_{4,2}(\alpha)}\propto \ket{\alpha}-\ket{i\alpha}+\ket{-\alpha}-\ket{-i\alpha}$, and $\ket{C_{4,3}(\alpha)}\propto \ket{\alpha}+i\ket{i\alpha}-\ket{-\alpha}-i\ket{-i\alpha}$ are four-headed cat states with different relative phases.

The generalized cat states (\ref{eq:pns1}) can be rewritten again by the number basis as
\begin{equation}
\ket{C_{d,k}(\alpha)}=\frac{de^{-|\alpha|^2/2}}{\sqrt{{\cal M}_{d,k}(\alpha)}}\sum^{\infty}_{n\in \mathbb{S}_k}\frac{\alpha^{n}}{\sqrt{n!}}\ket{n},
\end{equation}
where $\mathbb{S}_k=\{n| k\equiv n(\rm{mod~}d)\}$ \cite{psp1,psp2,Jkim2015} and ${\cal M}_{d,k}(\alpha)=\sum_{q,q'=0}^{d-1} \omega^{k(q'-q)} \bracket{\alpha\omega^{q'}}{\alpha\omega^{q}}= d^2 e^{-|\alpha|^2}\sum^{\infty}_{n\in \mathbb{S}_k}|\alpha|^{2n}/n!$. Thus, two states $\ket{C_{d,k}(\alpha)}$ and $\ket{C_{d,k'}(\alpha)}$ which have the same $d$ but different $k$, are orthonormal to each other, i.e.~$\bracket{C_{d,k}(\alpha)}{C_{d,k'}(\alpha)}=\delta_{k,k'}$. Conversely, each coherent state $\ket{\alpha\omega^q}$ can be represented by a superposition of the cat states $\ket{C_{d,k}(\alpha)}$ with $k\in\{0,1,...,d-1\}$, 
\begin{equation}
\label{eq:conE}
\ket{\alpha\omega^q}=\frac{1}{\sqrt{d}}\sum^{d-1}_{k=0}\omega^{kq}\sqrt{\frac{{\cal M}_{d,k}(\alpha)}{d}}\ket{C_{d,k}(\alpha)}.
\end{equation}
Note that $\ket{C_{d,k}(\alpha)}$ becomes closer to the ideal number state $\ket{k}$, as either $\alpha$ decreases or $d$ increases \cite{SWL19}. Thus, the generalized multi-component cat state in Eq.~(\ref{eq:pns1}) can be also referred to as the {\em pseudo number state} \cite{Jkim2015}. The fidelity between $\ket{C_{d,k}(\alpha)}$ and the ideal number state $\ket{k}$ is obtained as ${\cal F}(\ket{k},\ket{C_{d,k}(\alpha)})=|d e^{-\alpha}\alpha^{k}/\sqrt{k!{\cal M}_{d,k}(\alpha)}|^2$. We can observe that ${\cal F}(\ket{k},\ket{C_{d,k}(\alpha)})\rightarrow 1$ in the limit either $\alpha\rightarrow0$ with finite $d$ or $d\rightarrow\infty$ with finite $\alpha$ \cite{SWL19}. 

\section{Optimal phase estimation}
\label{sec:OPS}

\begin{figure}
\centering
\includegraphics[width=0.6\linewidth]{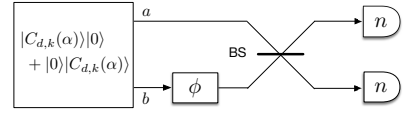}
\caption{Phase estimation setup. An entangled state of multi-component cat state $\ket{C_{d,k}(\alpha)}\ket{0}+\ket{0}\ket{C_{d,k}(\alpha)}$ is prepared and enter into two arms of an interferometer (denoted by mode $a$ and $b$). A phase shift $\phi$ is applied to one of the paths (mode $b$) of the interferometer. After combining the probe state in a $50:50$ beam splitter(BS), we measure photon number on both output modes.
In a lossless scenario, the ultimate bound can be achieved by photon number detection.
}\label{fig:fig1}
\end{figure}

Let us investigate the phase estimation with the generalized multi-component cat states. We are interested in the path-symmetric entangled states in the form of 
\begin{eqnarray}
\ket{\Psi_{d,k}(\alpha)}\equiv {\cal N}\left(\ket{C_{d,k}(\alpha)}\ket{0}+\ket{0}\ket{C_{d,k}(\alpha)}\right),
\end{eqnarray}
with the normalization factor $\cal N$. The phase estimation by using $\ket{\Psi_{2,0}(\alpha)}$ and $\ket{\Psi_{d,0}(\alpha)}$ has been investigated in Ref.~\cite{Joo11} and \cite{SYLee15}, respectively, which was shown to outperform the phase estimation with the NOON states under the same energy constraint. We here further consider the generalized form of $\ket{\Psi_{d,k}(\alpha)}$ with multi-component cat states to find out its advantage in quantum phase estimation by optimizing $d$ and $k$.

The sensitivity of phase estimation can be investigated by considering an interferometer as illustrated in Fig.~\ref{fig:fig1}. Assume that the generalized entangled coherent states $\ket{\Psi_{d,k}(\alpha)}$ enter the input of the interferometer. A phase shift operation $e^{i \phi \hat{b}^\dag\hat{b}}$ is applied to one of the paths in the interferometer. After combining by a 50:50 beam splitter, photon number measurements are performed on both output modes. In a lossless scenario, we can attain the ultimate bound by photon number detection \cite{SYL16}. We then estimate the phase difference $\phi$ between the two paths, aiming to achieve the sensitivity beating the SQL. For a single-shot measurement, the phase-estimation error is lower bounded by the inverse of the QFI, $\delta \phi \geq 1/\sqrt{F_Q}$, where $F_Q$ is the quantum Fisher information\cite{QFI}. Thus, the quality of phase estimation can be assessed through the quantum Fisher information.

After applying the phase shift operation $\ket{\Psi_{\text out}}=I\otimes e^{i \phi \hat{b}^\dag\hat{b}} \ket{\Psi_{d,k}(\alpha)}$, the quantum Fisher information can be obtained by
\begin{equation}
\begin{aligned}
\label{eq:QFI}
F_Q&=4\left(\bracket{\Psi'_{\text out}}{\Psi'_{\text out}}-|\bracket{\Psi'_{\text out}}{\Psi_{\text out}}|^2\right)\\
&=4\left(\bra{\Psi_{d,k}(\alpha)}\hat{n}^2_b\ket{\Psi_{d,k}(\alpha)}-\bra{\Psi_{d,k}(\alpha)}\hat{n}_b\ket{\Psi_{d,k}(\alpha)}^2\right),
\end{aligned}
\end{equation}
where $\ket{\Psi'_{\text out}}=\partial \ket{\Psi_{\text out}}/\partial\phi$ and $\hat{n}_b=\hat{b}^\dag\hat{b}$. We can evaluate 
\begin{eqnarray}
\bra{\Psi_{d,k}(\alpha)}\hat{n}^2_b&&\ket{\Psi_{d,k}(\alpha)}=\frac{{\cal N}^2}{{\cal M}_{d,k}(\alpha)}\sum^{d-1}_{q,q'=0}\omega^{k(q'-q)}\\
\nonumber
&&\times (|\alpha|^2\omega^{q-q'}+|\alpha|^4\omega^{2(q-q')})e^{(\omega^{q-q'}-1)|\alpha|^2},\\
\bra{\Psi_{d,k}(\alpha)}\hat{n}_b&&\ket{\Psi_{d,k}(\alpha)}=\frac{{\cal N}^2}{{\cal M}_{d,k}(\alpha)}\sum^{d-1}_{q,q'=0}\omega^{k(q'-q)}\\
\nonumber
&&\hspace{20mm}\times |\alpha|^2\omega^{q-q'}e^{(\omega^{q-q'}-1)|\alpha|^2}.
\end{eqnarray}

\begin{figure*}
\centering
\includegraphics[width=0.9\linewidth]{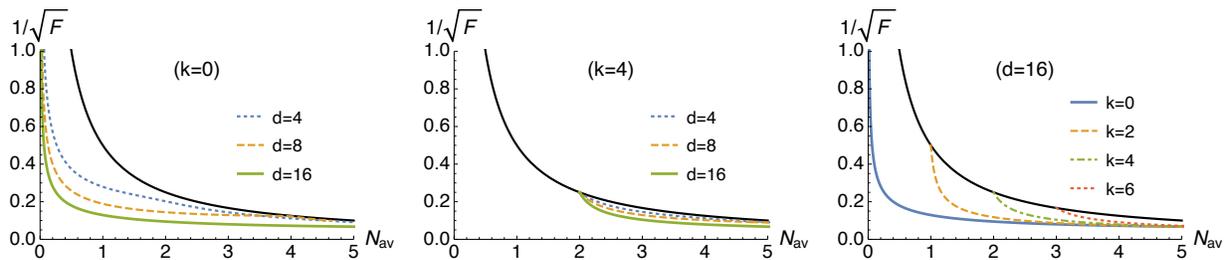}
\caption{Plots of the optimal phase estimation $1/\sqrt{F}$ against the average photon number $N_{av}$ using the entanglement of the generalized multi-component cat states, $\ket{\Psi_{d,k}(\alpha)}\sim \ket{C_{d,k}(\alpha)}\ket{0}+\ket{0}\ket{C_{d,k}(\alpha)}$, with different $k$ and $d$. The results are compared with the $1/\sqrt{F}$ obtained using the NOON states (black solid).
}\label{fig:fig2}
\end{figure*}

We plot the optimal phase estimation by the entangled coherent states $\ket{\Psi_{d,k}(\alpha)}$ with different $d$ and $k$ against the average photon number $N_{av}$ in Fig.~\ref{fig:fig2}. The average photon number for input mode is obtained here by $N_{av}=\bra{\Psi_{d,k}(\alpha)}\hat{n}_b\ket{\Psi_{d,k}(\alpha)}=\bra{\Psi_{d,k}(\alpha)}\hat{n}_a\ket{\Psi_{d,k}(\alpha)}$. The results are compared with the NOON state, $\ket{\Psi_{\rm NOON}}=(\ket{k}\ket{0}+\ket{0}\ket{k})\sqrt{2}$. The quantum Fisher information of $\ket{\Psi_{\rm NOON}}$ is $F_Q=k^2$ as $\bra{\Psi_{\rm NOON}} \hat{n}^2_b \ket{\Psi_{\rm NOON}}=k^2/2$ and $\bra{\Psi_{\rm NOON}}  \hat{n}_b \ket{\Psi_{\rm NOON}}=k/2$. We can observe that the phase estimation by the generalized cat states outperforms the optimal phase estimation by the NOON states in the regime,
\begin{equation}
\frac{k}{2}\leq N_{av}.
\end{equation}
When $k=0$, one can reproduce the results in Ref.~\cite{Joo11} for $d=2$ and in Ref.~\cite{SYLee15} for higher $d$. In the limit of decreasing the average photon number $N_{av}$, it approaches to the optimal phase estimation with the NOON states. The optimal phase estimation by $\ket{\Psi_{d,k}(\alpha)}$ becomes equivalent with the one by the NOON state, $\ket{\Psi_{\rm NOON}}$, at $N_{av}=k/2$ as shown in Fig.~\ref{fig:fig2}. Note that we can analytically verify this since $\ket{C_{d,k}(\alpha)}$ gets closer to the ideal number state $\ket{k}$ by decreasing $\alpha$ such that
\begin{equation}
\ket{\Psi_{d,k}(\alpha)}\rightarrow \frac{1}{\sqrt{2}}\big(\ket{k}\ket{0}+\ket{0}\ket{k}\big).
\end{equation}
Therefore, the optimal state for the phase estimation has the form of 
\begin{equation}
\ket{\Psi_{d,0}(\alpha)}={\cal N}\left(\ket{C_{d,0}(\alpha)}\ket{0}+\ket{0}\ket{C_{d,0}(\alpha)}\right)
\end{equation}
among all possible $\ket{\Psi_{d,k}(\alpha)}$ states. We can observe that the phase estimation performance can be enhanced further by increasing $d$ in many parts of the $N_{av}$ regime in Fig.~\ref{fig:fig2}. 

\begin{figure}
\centering
\includegraphics[width=0.99\linewidth]{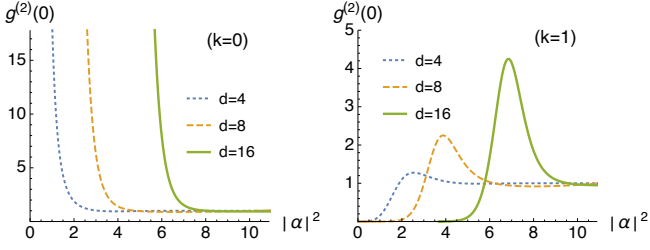}
\caption{The second-order correlation function $g^{(2)}(0)$ for multi-component cat states, $\ket{C_{d,0}(\alpha)}$ and $\ket{C_{d,1}(\alpha)}$, against $|\alpha|^2$ with $d=4,8,16$.
}\label{fig:fig3}
\end{figure}

The quantum Fisher information of Eq.~(\ref{eq:QFI}) can be represented by  
\begin{equation}
\label{eq:byg2}
F_Q=4{\cal N}^2 \langle \hat{n} \rangle \left\{ \big(g^{(2)}(0)-{\cal N}^2\big)\langle \hat{n} \rangle +1\right\}
\end{equation}
in terms of $\langle \hat{n} \rangle= \bra{C_{d,k}(\alpha)} \hat{n} \ket{C_{d,k}(\alpha)}$ and the second-order correlation function $g^{(2)}(0)=\langle \hat{a}^{\dag}\hat{a}^{\dag}\hat{a}\hat{a}\rangle/\langle \hat{a}^{\dag}\hat{a} \rangle^2$ of the generalized multi-component cat states $\ket{C_{d,k}(\alpha)}$. It can be also rewritten in terms of the Mandel-Q factor as $Q_M=\langle \hat{n} \rangle(g^{(2)}(0)-1)$  \cite{Sahota15,SYL16}. Note that the statistics with $g^{(2)}(0)<1$, $g^{(2)}(0)=1$, and $g^{(2)}(0)>1$ are called sub-Poissonian, Poissonian, and super-Poissonian, respectively. It shows that the higher $g^{(2)}(0)$ can lead to the lower quantum Cram\'er-Rao bound for a fixed amount of input energy. It was pointed out in Ref.~\cite{SYL16} that the super-Poissonianity of the single mode component of path-symmetric entangled state can enhance further the performance of the phase estimation. In Fig.~\ref{fig:fig3}, the second-order correlation functions $g^{(2)}(0)$ of $\ket{C_{d,0}(\alpha)}$ and $\ket{C_{d,1}(\alpha)}$ are plotted by changing the amplitude $\alpha$ with different $d$. It shows that $g^{(2)}(0)$ becomes larger as $d$ increases in many parts of the region. We can also observe that the second-order correlation functions for $k=0$ are much larger than $k=1$. Therefore, our result clearly demonstrates that the super-Poissonianity of $\ket{C_{d,k}(\alpha)}$ enhances the performance of the phase estimation. As a result, the optimal condition of $\ket{C_{d,k}(\alpha)}$ for the phase estimation in a noiseless scenario turns out to be $k=0$ with an arbitrarily high $d$.

Since QFI represents how sensitively we can detect a change of the phase parameter $\phi$, we might have to address the QFI about whether it may present different results depending on a configuration of a phase shifting operator. By averaging a two-mode input state over a common phase $\theta$ with an additional reference light \cite{Jarzyna12}, we obtain the phase average state
\begin{equation}
\rho_{d,k}(\alpha)=4{\cal N}^2\sum^{\infty}_{n\in \mathbb{S}_k}P_n\ket{\Psi_{\rm NOON}}\bra{\Psi_{\rm NOON}},
\end{equation}
where $\mathbb{S}_k=\{n| k\equiv n(\rm{mod~}d)\}$ and $P_n =d^2e^{-|\alpha|^2}|\alpha|^{2n}/n!{\cal M}_{d,k}(\alpha)$. Then, the corresponding QFI is the same as (\ref{eq:byg2}), regardless of a phase shifting operator $e^{i\phi\hat{n}_b}$ and $e^{-i\phi(\hat{n}_a-\hat{n}_b)/2}$. For a mixed state, the QFI is also obtained by the diagonalization of the mixed state.

\section{Effect of losses}
\label{sec:LOSS}

In this section, we analyze the effect of photon losses. The evolution of a quantum state under photon losses can be generally evaluated by solving the master equation \cite{Phoenix90}, $d\rho/dt=\gamma\hat{a}\rho\hat{a}^{\dag}-\gamma(\hat{a}^{\dag}\hat{a}\rho+\rho\hat{a}^{\dag}\hat{a})/2$, where $\hat{a} (\hat{a}^{\dag})$ is the annihilation (creation) operator, $\gamma$ is the decay constant, and $\eta=e^{-\gamma t}$ is the transmission rate under loss. The generalized cat states $\ket{C_{d,k}(\alpha)}$ evolve under losses to (normalization factor is omitted),
\begin{equation}
\label{eq:loss1}
\sum^{d-1}_{q,q'=0}\omega^{k(q'-q)}e^{(\omega^{q-q'}-1)|\alpha|^2(1-\eta)}\big|\alpha\sqrt{\eta}\omega^q\big\rangle \big\langle \alpha\sqrt{\eta}\omega^{q'}\big|.
\end{equation}
If we assume that the loss is weak ($\eta \lesssim 1$) under a limited energy constraint, i.e., $\alpha\sqrt{1-\eta}$ is small, the state in (\ref{eq:loss1}) can be written by 
\begin{equation}
\label{eq:loss2}
\begin{aligned}
&\big\{1-|\alpha|^2(1-\eta)\big\}{\cal M}_{d,k}(\alpha\sqrt{\eta})\big|C_{d,k}(\alpha\sqrt{\eta})\big\rangle \big\langle C_{d,k}(\alpha\sqrt{\eta})\big|\\
&\hspace{10mm}+|\alpha|^2(1-\eta){\cal M}_{d,k-1}(\alpha\sqrt{\eta})\big|C_{d,k-1}(\alpha\sqrt{\eta})\big\rangle \big\langle C_{d,k-1}(\alpha\sqrt{\eta})\big|.
\end{aligned}
\end{equation} 

In the interferometer for phase estimation, it is assumed that loss occurs after applying the phase shift operation in one mode, i.e., on the state $\ket{\Psi_{\text out}}=I\otimes e^{i \phi \hat{b}^\dag\hat{b}} \ket{\Psi_{d,k}(\alpha)}$. We can then evaluate its evolution under a weak loss as
\begin{equation}
\begin{aligned}
I\otimes e^{i \phi \hat{b}^\dag\hat{b}} \ket{\Psi_{d,k}(\alpha)} \xrightarrow[]{\eta}\rho_{d,k,\eta}(\alpha)
\end{aligned}
\end{equation}
where the detailed form of $\rho_{d,k,\eta}(\alpha)$ is given in Appendix~A. We here assume that both modes experience the same loss rate $\eta$. 

The quantum Fisher information for a mixed state $\rho$ can be obtained as \cite{mixedFish1,mixedFish2,mixedFish3}
\begin{equation}
\label{eq:FishMix}
\begin{aligned}
F_q=4\sum_i \lambda_i \left(\bracket{\lambda'_i}{\lambda'_i}-|\bracket{\lambda'_i}{\lambda_i}|^2\right) -\sum_{i \neq j} \frac{8\lambda_i\lambda_j}{\lambda_i+\lambda_j}\left|\bracket{\lambda'_i}{\lambda_j}\right|^2,
\end{aligned}
\end{equation}
where $\ket{\lambda_i}$ is the eigenvectors of $\rho$ with eigenvalues $\lambda_i$ s.t.~$\rho=\sum_i \lambda_i \ket{\lambda_i}\bra{\lambda_i}$, and $\ket{\lambda'_i}=\partial \ket{\lambda_i}/\partial \phi$. We thus diagonalize $\rho_{d,k,\eta}(\alpha)$ by its eigenvalues and eigenvectors as
\begin{equation}
\rho_{d,k,\eta}(\alpha)=\sum_{i=1}^4\lambda_i\ket{\lambda_i}\bra{\lambda_i},
\end{equation}
where the eigenvectors are given as (normalization factors are omitted)
\begin{equation}
\begin{aligned}
\ket{\lambda_1}&=\big|C_{d,k}(\alpha\sqrt{\eta})\big\rangle\big|0\big\rangle+\big|0\big\rangle\big|C_{d,k,\phi}(\alpha\sqrt{\eta})\big\rangle\\
\ket{\lambda_2}&=\big|C_{d,k}(\alpha\sqrt{\eta})\big\rangle\big|0\big\rangle-\big|0\big\rangle\big|C_{d,k,\phi}(\alpha\sqrt{\eta})\big\rangle\\
\ket{\lambda_3}&=\big|C_{d,k-1}(\alpha\sqrt{\eta})\big\rangle\big|0\big\rangle+\big|0\big\rangle\big|C_{d,k-1,\phi}(\alpha\sqrt{\eta})\big\rangle\\
\ket{\lambda_4}&=\big|C_{d,k-1}(\alpha\sqrt{\eta})\big\rangle\big|0\big\rangle-\big|0\big\rangle\big|C_{d,k-1,\phi}(\alpha\sqrt{\eta})\big\rangle
\end{aligned}
\end{equation}
where 
\begin{equation}
\big|C_{d,k,\phi}(\alpha)\big\rangle\equiv\frac{1}{\sqrt{{\cal M}_{d,k}(\alpha)}}\sum^{d-1}_{q=0}\omega^{-kq}\big|\alpha\omega^{q}e^{i\phi}\big\rangle,
\end{equation}
and each corresponding eigenvalues $\lambda_i$ are given in Appendix~B. By Eq.~(\ref{eq:FishMix}), we can thus calculate the quantum Fisher information of the state $\rho_{d,k,\eta}(\alpha)$. 

\begin{figure*}
\centering
\includegraphics[width=0.65\linewidth]{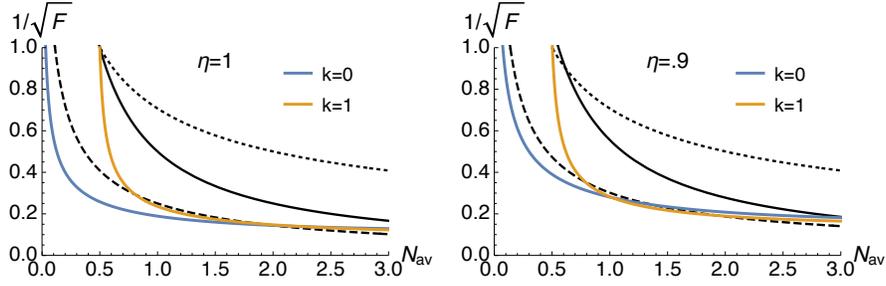}
\caption{Plots of the optimal phase estimation using $\ket{\Psi_{8,0}(\alpha)}$ and $\ket{\Psi_{8,1}(\alpha)}$ when $\eta=1$ (without loss) and $\eta=0.9$ (under loss), compared to the optimal phase estimation using the NOON (black solid curve) and TMSV state (black dashed curve) on the same loss rate $\eta$. As a reference, we plot the SQL (black dotted curve).
}\label{fig:fig4}
\end{figure*}

Based on the result presented in Sec.~\ref{sec:OPS}, we are particularly interested in the phase estimation with the optimal state, $\ket{\Psi_{d,0}(\alpha)}={\cal N}(\ket{C_{d,0}(\alpha)}\ket{0}+\ket{0}\ket{C_{d,0}(\alpha)})$, under the effect of photon losses. We additionally consider $\ket{\Psi_{d,1}(\alpha)}={\cal N}(\ket{C_{d,1}(\alpha)}\ket{0}+\ket{0}\ket{C_{d,1}(\alpha)})$ for comparison. Note that $\ket{C_{d,1}(\alpha)}$ evolves under losses to
\begin{equation}
\label{eq:Cd1loss}
\begin{aligned}
\ket{C_{d,1}(\alpha)}&\xrightarrow[]{\eta}A \big|C_{d,1}(\alpha\sqrt{\eta})\big\rangle \big\langle C_{d,1}(\alpha\sqrt{\eta})\big|\\
&\hspace{10mm}+B \big|C_{d,0}(\alpha\sqrt{\eta})\big\rangle \big\langle C_{d,0}(\alpha\sqrt{\eta})\big|,
\end{aligned}
\end{equation} 
where $A=\big\{1-|\alpha|^2(1-\eta)\big\}{\cal M}_{d,1}(\alpha\sqrt{\eta})$ and $B=|\alpha|^2(1-\eta){\cal M}_{d,0}(\alpha\sqrt{\eta})$, in which the proportion of $\ket{C_{d,0}(\alpha\sqrt{\eta})}$ increases as $\alpha$ increases for a given $\eta$. 

In Fig.~\ref{fig:fig4}, we plot the optimal phase estimations with and without the effect of loss by $\ket{\Psi_{d,0}(\alpha)}$ and $\ket{\Psi_{d,1}(\alpha)}$. We also compare these with the performance of the phase estimation using the NOON and the two-mode squeezed vacuum (TMSV) states \cite{Anisimov10}. It shows that the phase estimation by either $\ket{\Psi_{d,0}(\alpha)}$ or $\ket{\Psi_{d,1}(\alpha)}$ is better than the one obtained by the NOON state even in the presence of loss, while beating the SQL. In the region of low average photon number $N_{av}$, the phase estimation with the optimal state $\ket{\Psi_{d,0}(\alpha)}$ can also outperform the TMSV state. By increasing $N_{av}$, a slight crossover is observed between the maximum performance of $\ket{\Psi_{d,0}(\alpha)}$ and $\ket{\Psi_{d,1}(\alpha)}$ in the presence of loss. This is due to that the weight of $\ket{C_{d,0}(\alpha\sqrt{\eta})}$ in Eq.~(\ref{eq:Cd1loss}) grows with $N_{av}$. Note that the super-Poissonianity of $\ket{C_{d,0}(\alpha\sqrt{\eta}))}$ is larger than $\ket{C_{d,1}(\alpha\sqrt{\eta})}$ for an arbitrary $d$. 


\section{Generation scheme}
\label{sec:GS}
\begin{figure}[b]
\centering
\includegraphics[width=0.99\linewidth]{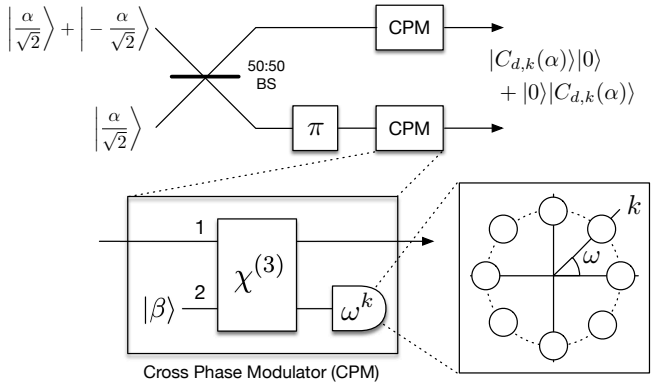}
\caption{A scheme to generate entangled multi-component cat states $\ket{C_{d,k}(\alpha)}\ket{0}+\ket{0}\ket{C_{d,k}(\alpha)}$ by employing a cross-phase modulator (CPM) based on cross-Kerr nonlinear interaction. In CPM, a heterodyne measurement is performed on mode 2 to identify $\ket{C_{d,k}(\alpha)}$ in the output mode 1.
}\label{fig:fig5}
\end{figure}

Let us consider a generation scheme of entangled multi-component cat states, $\ket{\Psi_{d,k}(\alpha)}\sim \ket{C_{d,k}(\alpha)}\ket{0}+\ket{0}\ket{C_{d,k}(\alpha)}$, using cross-phase modulators (CPMs) as illustrated in Fig.~\ref{fig:fig5}. A CPM can be implemented based on a cross-Kerr nonlinearity. Several schemes have been considered to produce multi-component cat states in cavity QED \cite{cavity}, circuit QED \cite{cQED,cQED1,Wang16}, and optomechanical systems \cite{opto}. Recently, a scheme to generate traveling optical multi-component cat state was proposed using Rb atoms confined in a hollow-core photonic crystal fiber (HC-PCF) \cite{SWL19}. Atomic vapor filling in HC-PCF has been studied as a platform to implement a cross phase shift operation \cite{Venkataraman12,Perrella13}, all-optical switches \cite{Venkataraman11,Bajcsy09}, and quantum memories \cite{Sprague14}. A conditional generation scheme of $\ket{C_{d,k}(\alpha)}$ of arbitrary $d$ and $k$ by CPM is introduced in Ref.~\cite{SWL19}.

Suppose that the a cat state $\ket{\alpha/\sqrt{2}}+\ket{-\alpha/\sqrt{2}}$ and coherent state $\ket{\alpha/\sqrt{2}}$ enter the input modes of a 50:50 beam splitter, and a $\pi$ phase shifter ($\ket{\alpha} \leftrightarrow \ket{-\alpha}$) is applied on the one output mode of the beam splitter. The output state is then given by
\begin{equation}
\begin{aligned}
\bigg(\left|\frac{\alpha}{\sqrt{2}}\right\rangle+\left|-\frac{\alpha}{\sqrt{2}}\right\rangle\bigg)\left|\frac{\alpha}{\sqrt{2}}\right\rangle&\xrightarrow{BS}\ket{\alpha}\ket{0}+\ket{0}\ket{-\alpha}\\
&\xrightarrow{I\otimes\pi}\ket{\alpha}\ket{0}+\ket{0}\ket{\alpha}.
\end{aligned}
\end{equation}

We then apply CPM on each mode, which is based on a cross-Kerr nonlinearity with interaction Hamiltonian $-\hbar\chi^{(3)}\hat{n}_1\hat{n}_2$ for time $t$ with the number operator $\hat{n}_i$ in $i$th mode. We here set $d \equiv 2\pi/\chi^{(3)} t$ as an integer $\geq 2$. Suppose that $\ket{\alpha}_1\ket{\beta}_2$ go through the two modes of CPM, where $\beta$ is assumed to be large s.t.~$\beta \gtrsim d$. By Eq.~(\ref{eq:conE}), the state in output modes can be then written by 
\begin{eqnarray}
\label{eq:genPSP}
\nonumber
&&e^{\frac{2\pi i}{d}\hat{n}_1\hat{n}_2}\ket{\alpha}_1\ket{\beta}_2\\
\nonumber
&=&\frac{e^{\frac{2\pi i}{d}\hat{n}_1\hat{n}_2}}{d}\Big(\sum^{d-1}_{k=0}\sqrt{\frac{{\cal M}_{d,k}(\alpha)}{d}}\ket{C_{d,k}(\alpha)}\Big)_1\Big(\sum^{d-1}_{j=0}\ket{C_{d,j}(\beta)}\Big)_2\\
\nonumber
&=&\frac{1}{\sqrt{d}}\sum^{d-1}_{k=0}\sqrt{\frac{{\cal M}_{d,k}(\alpha)}{d}}\ket{C_{d,k}(\alpha)}_1\Big(\frac{1}{\sqrt{d}}\sum^{d-1}_{j=0}\omega^{kj}\ket{C_{d,j}(\beta)}\Big)_2\\
&=&\sum^{d-1}_{j=0}\frac{\sqrt{{\cal M}_{d,k}(\alpha)}}{d}\ket{C_{d,k}(\alpha)}_1\ket{\beta\omega^k}_2.
\end{eqnarray}
Since $\beta> d$, the overlap between $\ket{\beta\omega^k}$ with different $k$ becomes negligible. We can perform heterodyne measurements to discriminate $k\in\{0,1,...,d-1\}$ in mode 1 in order to identify $\ket{C_{d,k}(\alpha)}$ in mode 1. Therefore, by applying CPM on each output modes as illustrated in Fig.~\ref{fig:fig5}, we can probabilistically generate the entangled multi-component cat states, 
\begin{equation}
\ket{\alpha}\ket{0}+\ket{0}\ket{\alpha}\xrightarrow{\rm CPM} \ket{C_{d,k}(\alpha)}\ket{0}+\ket{0}\ket{C_{d,k}(\alpha)}
\end{equation}
in a path-symmetric form.

\section{Conclusion}
\label{sec:CON}

We have investigated the performance of the phase estimation using generalized multi-component cat states which are the equally superposed coherent states on a circle, $\ket{C_{d,k}(\alpha)}\sim \sum^{d-1}_{q=0}\omega^{-kq}\ket{\alpha\omega^{q}}$ with $\omega=\exp{(2\pi i /d)}$. We showed that entangled multi-component cat states generally outperform the NOON and coherent states for estimating a phase in an interferometry. The optimal condition of the generalized multi-component cat states turns out to be $k=0$ regardless of the average photon number in a lossless scenario. It was also shown that the performance is enhanced further by increasing $d$ as the super-Poissonianity of the multi-component cat state increases.

We have also analyzed the effect of photon losses. We demonstrated that, in a low photon loss rate ($\leq 10 \%$), the phase estimation with entangled multi-component cat states beats the NOON and TMSV states in the region of a small energy constraint while beating the SQL. Notably, the optimal condition is shifted under the loss rate from $k=0$ to $k=1$ with the increasing average photon number. This is because $\ket{C_{d,1}(\alpha)}$ is changed to a state in the form of $\rho \sim  A \ket{C_{d,0}(\alpha)}\bra{C_{d,0}(\alpha)}+ B\ket{C_{d,1}(\alpha)}\bra{C_{d,1}(\alpha)}$ under losses as a mixture of $\ket{C_{d,0}(\alpha)}$ and $\ket{C_{d,1}(\alpha)}$, and the weight of $\ket{C_{d,0}(\alpha)}$ becomes larger as increasing the average photon number. Finally, we have proposed a scheme for producing entangled multi-component cat states by employing a cross-phase modulator, which may be feasible within current or near-term optical technologies.

An interesting path for further research may be the estimation of multiple phases \cite{Peter13,Liu16} with multi-component cat states. The performance of multi-phase estimation with the NOON \cite{Peter13} and the generalized entangled coherent states \cite{Liu16} have been studied. As the properties of the Cram\'er-Rao bound and QFI for multi-phase estimation differ with the single-phase estimation \cite{Liu20}, it is crucial to analyze further the performance and the optimal condition of the generalized cat state $\ket{C_{d,k}(\alpha)}$ to estimate more than two phase simultaneously. It may be also valuable to study the application to the metrology with quantum error correction \cite{Dur14} based on the encoding scheme using multi-component cat states \cite{Bergmann16,Grimsmo20}.

\acknowledgments
S.W.L. and J.K. were supported by KIAS Advanced Research Program (QP029902 and CG014604). S.-Y.L. was supported by Defense Acquisition Program Administration and Agency for Defense Development (Quantum Standoff Sensing Defense-Specialized Project).

\appendix
\begin{widetext} 
\section*{Appendix A. Entangled coherent states under losses}

Under the effect of photon losses on two modes (with the same rate $\eta$), $\ket{\Psi_{\text out}}=I\otimes e^{i \phi \hat{b}^\dag\hat{b}} \ket{\Psi_{d,k}(\alpha)}$ evolves to
\begin{equation}
\begin{aligned}
\rho_{d,k,\eta}(\alpha)&=\frac{{\cal N}^2}{{\cal M}_{d,k}(\alpha)}\sum^{d-1}_{q,q'=0}\omega^{k(q'-q)}e^{(\omega^{q-q'}-1)|\alpha|^2(1-\eta)}\Big(\big|\alpha\sqrt{\eta}\omega^q\big\rangle \big\langle\alpha\sqrt{\eta}\omega^{q'}\big|\otimes\big|0\big\rangle \big\langle0\big|\\
&\hspace{62mm}+\big|0\big\rangle \big\langle0\big|\otimes\big|\alpha\sqrt{\eta}\omega^qe^{i\phi}\big\rangle \big\langle\alpha\sqrt{\eta}\omega^{q'}e^{i\phi}\big|\Big)\\
&\hspace{5mm}+\frac{{\cal N}^2}{{\cal M}_{d,k}(\alpha)}\sum^{d-1}_{q,q'=0}\omega^{k(q'-q)}e^{-|\alpha|^2(1-\eta)}\Big(\big|\alpha\sqrt{\eta}\omega^q\big\rangle \big\langle0\big| \otimes \big|0\big\rangle\big\langle\alpha\sqrt{\eta}\omega^{q'}e^{i\phi}\big|\\
&\hspace{58mm}+\big|0\big\rangle\bra{\alpha\sqrt{\eta}\omega^{q'}}\otimes\ket{\alpha\sqrt{\eta}\omega^qe^{i\phi}}\big\langle0\big|\Big).
\end{aligned}
\end{equation}
Assume a weak loss with a limited energy constraint i.e., small $\alpha\sqrt{1-\eta}$, it can be written by
\begin{equation}
\begin{aligned}
\rho_{d,k,\eta}(\alpha)&=\frac{{\cal N}^2(1-|\alpha|^2(1-\eta))}{{\cal M}_{d,k}(\alpha)}\sum^{d-1}_{q,q'=0}\omega^{k(q'-q)}\Big(\big|\alpha\sqrt{\eta}\omega^q\big\rangle \big\langle\alpha\sqrt{\eta}\omega^{q'}\big|\otimes\big|0\big\rangle \big\langle0\big|\\
&\hspace{62mm}+\big|0\big\rangle \big\langle0\big|\otimes\big|\alpha\sqrt{\eta}\omega^qe^{i\phi}\big\rangle \big\langle\alpha\sqrt{\eta}\omega^{q'}e^{i\phi}\big|\Big)\\
&\hspace{5mm}+\frac{{\cal N}^2|\alpha|^2(1-\eta)}{{\cal M}_{d,k}(\alpha)}\sum^{d-1}_{q,q'=0}\omega^{(k-1)(q'-q)}\Big(\big|\alpha\sqrt{\eta}\omega^q\big\rangle \big\langle\alpha\sqrt{\eta}\omega^{q'}\big|\otimes\big|0\big\rangle \big\langle0\big|\\
&\hspace{62mm}+\big|0\big\rangle \big\langle0\big|\otimes\big|\alpha\sqrt{\eta}\omega^qe^{i\phi}\big\rangle \big\langle\alpha\sqrt{\eta}\omega^{q'}e^{i\phi}\big|\Big)\\
&\hspace{5mm}+\frac{{\cal N}^2e^{-|\alpha|^2(1-\eta)}}{{\cal M}_{d,k}(\alpha)}\sum^{d-1}_{q,q'=0}\omega^{k(q'-q)}\Big(\big|\alpha\sqrt{\eta}\omega^q\big\rangle \big\langle0\big| \otimes \big|0\big\rangle\big\langle\alpha\sqrt{\eta}\omega^{q'}e^{i\phi}\big|\\
&\hspace{58mm}+\big|0\big\rangle\bra{\alpha\sqrt{\eta}\omega^{q'}}\otimes\ket{\alpha\sqrt{\eta}\omega^qe^{i\phi}}\big\langle0\big|\Big).
\end{aligned}
\end{equation}

\section*{Appendix B. Eigenvalues of $\rho_{d,k,\eta}(\alpha)$}

The eigenvalues of $\rho_{d,k,\eta}(\alpha)$ can be evaluated as $\lambda_i=E_i/(E_1+E_2+E_3+E_4)$, where
\begin{equation}
\begin{aligned}
E_1&=\frac{{\cal M}_{d,k}(\alpha\sqrt{\eta})}{{\cal M}_{d,k}(\alpha)}\bigg(\frac{1+|\bracket{0}{C_{d,k}(\alpha\sqrt{\eta})}|^2}{1+|\bracket{0}{C_{d,k}(\alpha)}|^2}\bigg)\frac{1-|\alpha|^2(1-\eta)+e^{-|\alpha|^2(1-\eta)}}{2},\\
E_2&=\frac{{\cal M}_{d,k}(\alpha\sqrt{\eta})}{{\cal M}_{d,k}(\alpha)}\bigg(\frac{1-|\bracket{0}{C_{d,k}(\alpha\sqrt{\eta})}|^2}{1+|\bracket{0}{C_{d,k}(\alpha)}|^2}\bigg)\frac{1-|\alpha|^2(1-\eta)-e^{-|\alpha|^2(1-\eta)}}{2},\\
E_3&=\frac{{\cal M}_{d,k-1}(\alpha\sqrt{\eta})}{{\cal M}_{d,k}(\alpha)}\bigg(\frac{1+|\bracket{0}{C_{d,k-1}(\alpha\sqrt{\eta})}|^2}{1+|\bracket{0}{C_{d,k}(\alpha)}|^2}\bigg)\frac{|\alpha|^2(1-\eta)}{2},\\
E_4&=\frac{{\cal M}_{d,k-1}(\alpha\sqrt{\eta})}{{\cal M}_{d,k}(\alpha)}\bigg(\frac{1-|\bracket{0}{C_{d,k-1}(\alpha\sqrt{\eta})}|^2}{1+|\bracket{0}{C_{d,k}(\alpha)}|^2}\bigg)\frac{|\alpha|^2(1-\eta)}{2}.
\end{aligned}
\end{equation}

\end{widetext}

\end{document}